\documentclass[aps,prl,
twocolumn, 
amsfonts,amssymb,amsmath,
tightenlines,
twoside,
letterpaper,%a4paper,
bibnotes,
]{revtex4}

\usepackage{bm,color,mathrsfs,graphicx}

\begin{document}

\title{Koshino-Taylor effect in graphene}

\author{A. Cano}
\email{cano@ill.fr}
\affiliation{Institut Laue-Langevin, 6 rue Jules Horowitz, B.P. 156, 38042 Grenoble, France}

\date{\today}

\begin{abstract}
We discuss the phonon-assisted scattering of electrons by defects, i.e., the so-called Koshino-Taylor effect, in graphene. The two-dimensional character of graphene implies that the strength of the Koshino-Taylor effect can be considerably larger than in ordinary metals. 
We show that at finite temperatures the defect-induced resistivity formally diverges in the thermodynamic limit, having a non-analytic $T\ln T$ component when finite size effects are taken into account.
\end{abstract}


\maketitle

The possibility of fabricating free-standing graphene sheets has recently been demonstrated experimentally \cite{Meyer07,Bunch07}. In this form graphene combines its unusual electronic features with additional mechanical peculiarities, giving rise to new specific properties (see, e.g., Ref. \cite{CastroNeto08}). 
The resistivity of suspended graphene, for example, shows a clear dependence on temperature \cite{Du08,Bolotin08} in contrast to samples supported by a substrate where it has an important $T$-independent contribution \cite{Novoselov04}.

The most distinctive feature of graphene is the Dirac spectrum of its charge carriers \cite{CastroNeto08}. Suspended graphene, in its own turn, is inevitably subjected to both in-plane and out-of-plane distortions.
As shown in Ref. \cite{Mariani08}, this gives rise to two different contributions in its low-temperature resistivity. Whereas the scattering of electrons with in-plane phonons provides a $T^4$ contribution (see also \cite{Hwang08}), the scattering with out-of-plane ones gives an unusual $T^{5/2}\ln T$ behavior. On the other hand, the precise nature of the defects responsible for the residual (temperature-independent) resistivity is still unclear in graphene \cite{Peres06,CastroNeto08}. 
If they could be described in terms of Dirac-delta-function potentials, the resulting 
resistivity would not depend on the carrier density $n$. In the experiments, however, it turns out to be inversely proportional to $n$ \cite{Novoselov04}. 
This non-trivial behavior can be explained in different ways: due to unscreened Coulomb potentials \cite{Nomura06,Ando06}, frozen corrugations of the graphene sheet \cite{Katsnelson08}, the scattering involving the midgap states that may be created by disorder \cite{Stauber07}, etc.
The renormalization of the impurity scattering due to e.g. electron-electron interaction is another source of temperature-dependent contributions to the resistivity. Specifically in graphene, Friedel oscillations in the exchange field yield a linear-in-$T$ behavior (which is absent in the case of Coulomb scatterers) \cite{Cheianov06}. In this paper we consider the phonon-assisted scattering of electrons by defects, i.e., the so-called Koshino-Taylor effect \cite{Koshino60,Taylor62,Koshino63,Reizer87,Mahan}. This effect is operative for any kind of defect and, in the case of graphene, is expected to be unusually large due to the two-dimensional character of the system. As we show below, both in-plane and out-of-plane phonons can yield a $T \ln T$ behavior of the low-temperature resistivity in graphene through the Koshino-Taylor effect. 

The key point in Koshino-Taylor effect are the local fluctuations of the system as experienced by defects. In fact, in real systems, the position of the defects is not fixed but fluctuates, what eventually translates into a temperature dependence of the defect-induced resistivity \cite{Koshino60,Taylor62,Koshino63,Reizer87,Mahan}. 
In ordinary metals this a rather small effect that, nevertheless, can be verified experimentally after some effort \cite{Ptitsina97}. 
In the case of free-standing graphene, conversely, an unusually large effect can be anticipated because of the following. 
Within a first approximation, it can be assumed that the changes in the positions of the defects coincide with the lattice fluctuations at the corresponding points. 
Thus, the residual resistivity obtained under the assumption of static impurities is effectively dressed by a Debye-Waller-like factor due to these fluctuations. 
Local fluctuations in a two-dimensional (2D) lattice formally diverge and, in accordance with the Mermin-Wagner theorem, such a divergence should destroy the crystalline order. In practice, however, long-wavelength fluctuations are suppressed by finite size effects for example. This stabilizes the lattice, but local fluctuations may remain quite large. The Koshino-Taylor effect, benefiting from these large local fluctuations, is thus expected to be amplified in free-standing graphene sheets.

The simplest way to confirm this expectation is by means of the Boltzmann transport theory. 
This theory is applicable in the case of gated and/or doped graphene with a finite (and tunable) carrier density (i.e. the so-called extrinsic graphene) \cite{Auslender07,CastroNeto08}. 
Despite interference contributions are not captured within this approach, it suffices to reproduce correctly the qualitative behavior in our case \cite{Reizer87}.
The resistivity then can be written as
\begin{align}
\rho = {2\over e^2 v_F^2 \nu} {1\over \tau}, 
\label{}\end{align}
where $v_F$ is the Fermi velocity, $\nu$ is the density of states at the Fermi level, and $\tau$ is the transport relaxation time at the Fermi energy averaged over the angles (see, e.g., Ref. \cite{LifshitzPitaevskii}). In graphene $\nu = 2k_F /(\pi \hbar v_F)$, where the Fermi wavevector $k_F$ can be expressed in terms of the carrier concentration as $k_F = (\pi n)^{1/2}$.

The relaxation time $\tau $ is obtained from the so-called collision integral \cite{LifshitzPitaevskii}. So let us reconsider, first of all, the scattering of an electron due to a single defect originally situated at $\mathbf r = 0$. The scattering potential associated with the defect is $V (\mathbf r - \mathbf u)$, where $\mathbf u$ represents its displacement. Let $\mathbf k$ and $\mathbf k'$ be the initial and final wavevectors of the electron, and $|N\rangle $ and $| N' \rangle$ the initial and final phonon states respectively. The scattering rate then can be written as
\begin{align}
{2\pi \over \hbar }|V_{\mathbf k',\mathbf k }|^2 
|\langle N'| e^{i \mathbf K \cdot \mathbf u} | N \rangle |^2 \delta (E_f - E_i).
\label{scattering-rate}\end{align}
Here $V_{\mathbf k',\mathbf k }$ is the matrix element one would obtain by neglecting the displacement of the defect (i.e. taking $\mathbf u=0$), $\mathbf K = \mathbf k'- \mathbf k$ is the change in the wavevector of the electron, and $E_i$ ($E_f$) represents the total energy of the initial (final) state. It is worth noticing that the vector $\mathbf K$ lies on the graphene plane, so only the scattering with the in-plane modes turns out effective here.

In the case of small displacements, it suffices to retain the lowest order terms of expansion of the exponential in powers of $\mathbf u$: $e^{i \mathbf K \cdot \mathbf u} \simeq 1 + i\mathbf K \cdot \mathbf u - {1\over 2}(\mathbf K \cdot \mathbf u)^2$. Furthermore, by assuming that the displacement of the defect coincides with the displacement of the lattice, i.e., the defect is pinned to the lattice, the vector $\mathbf u$ can be written as a sum of creation $a_{\mathbf q,s} ^+$ and annihilation operators $a_{\mathbf q,s}$ for the phonon modes.
Thus the main contribution to the scattering rate \eqref{scattering-rate} is given by two types of processes. On one hand we have one-phonon processes, for which we have to consider the matrix elements $\langle N - 1| a | N \rangle = \langle N | a ^+| N - 1 \rangle = N^{1/2} $, where $N$ is the phonon distribution function. On the other hand we have processes involving no change in the phonon state, which are associated the matrix elements $\langle N | N \rangle  = 1$ and 
$\langle N | a a^+ | N \rangle = \langle N | a^+ a | N \rangle - 1= N$. 
Within our approximations the graphene sheet can be considered as a continuum elastic medium with the two in-plane modes having the same velocity.
Thus, after summation over the phonon states, the scattering rate for the electron is given by
\begin{widetext}
\begin{align}
{2\pi \over \hbar } |V_{\mathbf k',\mathbf k }|^2 
\Big\{
&
\Big[
1
-
\sum_{\mathbf q} 
{\hbar K^2 \over 2 \sigma \omega_{\mathbf q}}
(1+2 N_{\mathbf q})
\Big]
\delta(\varepsilon _{\mathbf k'} - \varepsilon _{\mathbf k} )
+ 
\sum_{\mathbf q} 
{\hbar K^2 \over 2 \sigma \omega_{\mathbf q}}
\big[
(1+N_{\mathbf q}) \delta(\varepsilon _{\mathbf k'} - \varepsilon _{\mathbf k} -\hbar \omega_{\mathbf q})
+
N_{\mathbf q} \delta(\varepsilon _{\mathbf k'} - \varepsilon _{\mathbf k} + \hbar \omega_{\mathbf q})
\big]
\Big\},
\label{scattering-rate-expanded}\end{align}
where $\sigma $ is the mass density and $\omega_{\mathbf q } = c q$ is the phonon frequency (here and hereafter the area of the system is taken as unity).

In the case of a system of electrons, this scattering rate has to be multiplied by the electron distribution function $f_{\mathbf k}$ for the initial state and by $1 - f_{\mathbf k'}$ for the final state. The collision integral then takes the form $I = I_0 + I_{el} + I_{in} $, where 
\begin{subequations}\begin{align}
I_0 &= {2\pi \over \hbar }n_i\sum _{\mathbf k'}
|V_{\mathbf k',\mathbf k }|^2 
(f_{\mathbf k'} - f_{\mathbf k})
\delta(\varepsilon _{\mathbf k'} - \varepsilon _{\mathbf k} ),
\label{ordinary-collission-integral}
\\
I_{el} &= - 
{2\pi \over \hbar } n_i
\sum _{\mathbf k',\mathbf q}
|V_{\mathbf k',\mathbf k }|^2 
{\hbar K^2 \over 2\sigma \omega_{\mathbf q}}
(f_{\mathbf k'} - f_{\mathbf k})
(1+2 N_{\mathbf q} )
\delta(\varepsilon _{\mathbf k'} - \varepsilon _{\mathbf k} )
,
\label{correction-elastic}\\
I_{in} &=  
{2\pi \over \hbar } n_i
\sum _{\mathbf k',\mathbf q}
|V_{\mathbf k',\mathbf k }|^2 
{\hbar K^2 \over 2 \sigma \omega_{\mathbf q}}
\big\{
[ f_{\mathbf k'} (1 - f_{\mathbf k})(1+N_{\mathbf q}) + f_{\mathbf k} (1 - f_{\mathbf k'})N_{\mathbf q}]
\delta(\varepsilon _{\mathbf k'} - \varepsilon _{\mathbf k} -\hbar \omega_{\mathbf q})
\nonumber \\
&
\qquad \qquad \qquad \qquad \qquad \qquad +
[ f_{\mathbf k'} (1 - f_{\mathbf k})N_{\mathbf q}+ f_{\mathbf k} (1 - f_{\mathbf k'})(1+N_{\mathbf q})]
\delta(\varepsilon _{\mathbf k'} - \varepsilon _{\mathbf k} + \hbar \omega_{\mathbf q})
\big\},
\label{correction-inelastic}
\end{align}\label{collision-integral}\end{subequations}
\end{widetext}
with $n_i$ being the density of defects. For the sake of concreteness, we assume in the following that the deviation from equilibrium is due to the presence of a static and spatially homogeneous electric field. If such a deviation is small, Eqs. \eqref{collision-integral} can be linearized with respect to $g =f - f_0 = - {\partial f_0 \over \partial \varepsilon }\varphi = {f _0 (1-f_0)\over T}\varphi$, where $f_0(\varepsilon ) = (e^{(\varepsilon - \mu)/T} + 1 )^{-1}$, assuming that the phonon distribution function takes its value of equilibrium: $N=(e^{\hbar \omega_{\mathbf q} /T}- 1)^{-1}$. 

The integral \eqref{ordinary-collission-integral} is nothing but the integral that one obtains neglecting the fluctuations in the positions of the defects. This further gives a residual resistivity $\rho_0$, as can be seen in standard textbooks (e.g. Ref. \cite{LifshitzPitaevskii}). As regards \eqref{correction-elastic}, it can be written as $I_{el} = - g / \tau_{el} $ after some standard manipulations (see e.g. Ref. \cite{LifshitzPitaevskii}). Here
\begin{align}
{1\over \tau_{el}}& = -
\sum _{\mathbf k',\mathbf q}
W(\theta)
{\hbar k^2_F \over \sigma \omega_{\mathbf q}}
[1 + 2N(\hbar \omega_{\mathbf q})]
\delta(\varepsilon _{\mathbf k'} - \varepsilon _{\mathbf k} ),
\label{tau-elastic}
\end{align}
where $W (\theta)= {2\pi \over \hbar } n_i |V_{\mathbf k',\mathbf k }|^2 
(1-\cos \theta )^2$ ($\theta$ being the angle between $\mathbf k'$ and $\mathbf k$). At $T=0$ this gives a correction to the residual resistivity obtained from the previous integral \eqref{ordinary-collission-integral}. This correction can be estimated as 
\begin{align}
\delta\rho_0 \sim - {\hbar^2 k_F^2 \over M \Theta}\rho_0,
\end{align}
where $M$ is the mass of the unit cell and $\Theta$ is the Debye temperature. In the usual case $\varepsilon_F=\hbar^2 k_F^2/(2m) $, where $m$ is the mass of the electron. In the above expression we then have ${\hbar^2 k_F^2 \over M \Theta} = {m\over M}{\varepsilon _F\over \Theta}$. This factor is typically $\sim 10^{-2}$ for ordinary metals. In the case of graphene  $\varepsilon_F = \hbar v_F k_F$, so the above factor can be written as ${\varepsilon_F\over M v_F^2}{\varepsilon_F\over \Theta}$. This factor can reach the same value than in ordinary metals for large enough concentrations of charge carriers (i.e., large values of the Fermi energy $\varepsilon_F$). Actually at such a concentrations deviations from the Dirac spectrum might be important. However, as we can see, the precise form of the spectrum is not the crucial point for having the above correction to the residual resistivity due to the Koshino-Taylor effect.

Let us now turn to the dependence on the temperature of the defect-induced resistivity. This dependence is due to both, the elastic processes giving rise to \eqref{tau-elastic}, and the inelastic processes taken into account in 
\begin{widetext}
\begin{align}
I_{in} & \simeq 
{2\pi \over \hbar }n_i 
\sum _{\mathbf k',\mathbf q}
|V_{\mathbf k',\mathbf k }|^2 
{\hbar K^2 \over 2\sigma \omega_{\mathbf q} T}
[f_{0}(\varepsilon _{\mathbf k'})- f_{0}(\varepsilon _{\mathbf k} )]N(\hbar \omega_{\mathbf q})[1 + N(\hbar \omega_{\mathbf q})] (\varphi_{\mathbf k'}-\varphi_{\mathbf k})
[
\delta(\varepsilon _{\mathbf k'} - \varepsilon _{\mathbf k} -\hbar \omega_{\mathbf q})
-
\delta(\varepsilon _{\mathbf k'} - \varepsilon _{\mathbf k} + \hbar \omega_{\mathbf q})
]
\label{correction-inelastic-lin}.
\end{align}
\end{widetext}
In fact, the remaining contribution at low temperatures is due to this later integral (see Ref. \cite{Koshino63}). This is basically because here we have the factors $N(1+N)$ whereas in \eqref{correction-elastic} we have $N$ only. As a result of these factors the differences in the electron energies $\varepsilon _{\mathbf k'} - \varepsilon _{\mathbf k} =\pm \hbar \omega_{\mathbf q}$ are small, so we can put
\begin{align}
f_{0}(\varepsilon _{\mathbf k'})- f_{0}(\varepsilon _{\mathbf k} ) \approx {\pm \hbar \omega_{\mathbf q} } \Big({\partial f_{0}(\varepsilon)\over \partial \varepsilon}\Big)_{\varepsilon =\varepsilon_{\mathbf k}}.
\label{}
\end{align}
The integral \eqref{correction-inelastic-lin} then takes the form $I_{in} \approx - g/ \tau _{in}$, where 
\begin{align}
{1\over \tau_{in}} & =
\sum _{\mathbf k',\mathbf q}
W(\theta)
{2 \hbar^2 k^2_F \over \sigma  T}
N(\hbar \omega_{\mathbf q})[1 + N(\hbar \omega_{\mathbf q})]
\delta(\varepsilon _{\mathbf k'} - \varepsilon _{\mathbf k}).
\label{correction-inelastic-lin-lowT}
\end{align}
The corresponding contribution to the resistivity can be estimated as \begin{align}
\delta \rho \sim { \hbar ^2 k_F^2 \over M \Theta }{T\over \Theta} \ln \Big( {T \over T_0}\Big)\rho_0,
\label{rho(T)KT}\end{align}
where $T_0$ is temperature associated with the infrared cutoff of the integral over the phonon wavevectors. In the expression obtained for an ordinary (3D) metal one has an additional factor $T/\Theta$ instead of $\ln (T/T_0)$ \cite{Koshino60,Taylor62}. This latter logarithmic factor is the fingerprint of the local fluctuations of the 2D lattice, whose largeness is ultimately determined by the quantity $T_0$ (see below). 

Let us now discuss the possibility of a similar $T \ln T$ contribution due to the out-of-plane distortions of the graphene sheet. The virtual changes in the form of the defect potentials owing to the (local) deformations of the lattice brings about this possibility. 
That is, changes described as $(1 + \lambda u_{ll}) V(\mathbf r)$, where $u_{ll}$ is the trace of the strain tensor and $\lambda $ is a constant. This effect was first discussed in Ref. \cite{Klemens63} for ordinary metals. In the case of graphene the strain tensor takes the from $u_{ij} = {1\over 2} [\partial_i u_j + \partial_j u_i + (\partial _i h)(\partial _j h)]$,
where $h$ characterizes the out-of-plane distortions \cite{Mariani08}. Here we see that local in-plane deformations $\langle u_{ll} \rangle$ are possible as a result of out-of-plane fluctuations of the lattice. The resistivity then can be written as $\rho \sim (1 + \lambda \langle u_{ll} \rangle )\rho_0$.  
In the absence of tension in the graphene sheet the dispersion law of the out-of-plane phonons is $\omega_{\mathbf q }^{(h)} = \kappa q^2 $. At low temperatures we then have
\begin{align}
\langle u_{ll} \rangle \approx { \hbar \over 2\pi M \kappa } 
\bigg[{1\over 2} +  {T\over \Theta^{(h)}} \ln \bigg({T \over T_0^{(h)}}\bigg)\bigg].    
\end{align} 
Here $\Theta^{(h)}$ is the Debye temperature for the out-of-plane phonons and $T_0^{(h)}$ is associated with the infrared cutoff for these phonons (see below). It is worth mentioning that if the graphene sheet is under tension, the dispersion law of the out-of-plane phonons becomes linear-in-$q$ \cite{CastroNeto08}. In this case the temperature dependence of $\langle u_{ll} \rangle$ changes from $T\ln T $ to $T^3$, and the corresponding contribution to the resistivity is then similar to that obtained from the mean value of the square of the local deformations $\langle u_{ll}^2 \rangle$ (which is due to in-plane distortions). This sensitivity to applied tension is absent in the contribution \eqref{rho(T)KT} due to in-plane fluctuations. It is worth noticing also that these two contributions differ in their dependence on the Fermi wavevector $k_F$, i.e., on the carrier density. In the out-of-plane contribution this dependence comes from the matrix elements of the (unchanged) scattering potential $V$ only. Consequently, its dependence on the carrier density turns out to be the same than the one of bare resistivity $\rho_0$. This is not the case of contribution due to in-plane phonons, where there is an additional dependence coming from the fluctuations [see Eq. \eqref{rho(T)KT}].

The dependence on the temperature of the resistivity is quite apparent in suspended graphene \cite{Du08,Bolotin08}. Far from the charge-neutrality point, this can be explained as due to electron-phonon scattering for temperatures above 50 K \cite{Bolotin08,footnote2}. At lower temperatures this scattering is expected to give $T^4$ and $T^{5/2}\ln T$ contributions due to in-plane and out-of-plane phonons respectively \cite{Mariani08}. So, if the temperature is low enough, the Koshino-Taylor effect can overwhelm the electron-phonon scattering giving rise to a $T\ln T$ behavior in the resistivity. The range of temperatures at which the Koshino-Taylor effect can be dominant is difficult to estimate realistically given its strong dependence on the nature of the defects, their concentration, wavevector cutoffs, etc. 
As regards this latter quantities, for example, a crude estimate can be obtained from the smallest possible wavevector in a system with a finite size $L$: $q_\text{min} \sim 2\pi / L$ \cite{CastroNeto08}. In the case of out-of-plane phonons, however, the combined effect of fluctuations and anharmonicity yields a strong renormalization of the bending rigidity $\kappa $ \cite{Mariani08,Aronovitz89,Fasolino07}. This eventually suppresses out-of-plane fluctuations, making it possible the low temperature flat phase of the sheet. This result can be seen as due to an anharmonicity-induced infrared cutoff \cite{Mariani08}, which is the relevant one for large enough samples \cite{Fasolino07} (i.e., in the thermodynamic limit). But in the case of in-plane phonons the tendency is just the contrary. In this case, nonlinear effects make the divergence of local in-plane fluctuations even stronger \cite{Aronovitz89}. In consequence, by employing Eq. \eqref{rho(T)KT} with $T_0 \sim \hbar c q_\text{min}$, one is actually underestimating the Koshino-Taylor effect if the size of the sample is large enough. 

At this point, it is worth mentioning that a similar dependence on the temperature is obtained from the renormalization of the impurity scattering due to the Friedel oscillations of the corresponding exchange field \cite{Cheianov06}. But in contrast to the Koshino-Taylor effect, such a renormalization is not operative in the case of Coulomb scatterers and does not scale with the size of the system. In any case, it is worth noticing there is a correlation between the temperature-dependent part of these contributions and the residual $T=0$ resistivity which might be useful in their experimental identification \cite{Ptitsina97}.

In summary, we have discussed the peculiarities of the Koshino-Taylor effect in graphene. Both in-plane displacements of the defects and deformations of the scattering potentials due to out-of-plane distortions of the graphene sheet yield a $T\ln T$ dependence of the defect-induced resistivity. These two contributions, however, differ in their dependence on external parameters such as the carrier concentration and the tension of the graphene sheet. The key ingredient in the Koshino-Taylor effect is the local fluctuations of the lattice which, in graphene, can be quite large due to the two-dimensional character of the system. Graphene thus represents a model system to study the Koshino-Taylor effect, an effect somewhat elusive in ordinary metals.

I acknowledge E. Bascones, M. Civelli, E. Kats, A. Levanyuk, S. Minyukov, I. Paul and R. Whitney for very fruitful discussions and Fundaci\' on Ram\' on Areces.


\begin{references}
\bibitem{footnote2} It is worth noticing that the linear-in-$T$ contribution observed in \cite{Bolotin08} shows a carrier-density dependence not captured by ordinary electron-phonon scattering. This dependence is compatible with the Koshino-Taylor contribution due to out-of-plane fluctuations and unscreened Coulomb scatterers.
\end{references}
\end{document}